\documentstyle[epsf]{l-aa}
\begin{document}
\def\HI{H{\sc i}\ }
\thesaurus{03(11.09.4; 11.05.2; 09.13.2; 11.19.2)}
\title{Star formation and the interstellar medium in low surface
brightness galaxies}
\subtitle{II. Deep  CO observations of
  low surface brightness disk galaxies}

\author{W.J.G.  de Blok\thanks{\emph{Present address:} School of
Physics, University of Melbourne, Parkville VIC 3052, Australia} and
J.M.  van der Hulst}

\institute{Kapteyn Astronomical Institute, P.O.~Box 800, NL-9700 AV
  Groningen, The Netherlands,\\ edeblok@physics.unimelb.edu.au, vdhulst@astro.rug.nl }
\offprints{W.J.G. de Blok}

\date{Received: ; accepted: }

\maketitle
\markboth{de Blok \& van der Hulst: Star formation \& ISM in LSB
  galaxies}{}
\begin{abstract}
  We present deep, pointed $^{12}$CO($J=2-1$) observations of three
  late-type LSB galaxies. The beam-size was small enough that we could
  probe different environments (\HI maximum, \HI mininum, star forming
  region) in these galaxies. No CO was found at any of the positions
  observed.  We argue that the implied lack of molecular gas is real
  and not caused by conversion factor effects. The virtual absence of
  a molecular phase may explain the very low star formation rates in
  these galaxies.
\keywords{Galaxies: ISM -- Galaxies: evolution -- ISM: molecules --
  Galaxies: spiral}
\end{abstract}

\section{Introduction}

Low Surface Brightness (LSB) galaxies exhibit some of the most extreme
properties known for disk galaxies.  This class itself can be split into
the group of disk dominated, late-type field galaxies with absolute
magnitudes between $M_B \sim -17$ and $M_B \sim -19$ and central surface
brightnesses $\mu_0(B) \simeq 23.5$ mag arcsec$^{-2}$, and the much
smaller group of giant LSB galaxies (``Malin-1 cousins'', Sprayberry et
al 1995 and Knezek 1993).  Judging from their morphology these latter
galaxies have undergone an evolutionary history quite different from the
late-type field LSB galaxies which we will consider in this paper. 

The extreme gas-richness (McGaugh \& de Blok 1997) and low metallicities
(McGaugh 1994) of the late-type LSB galaxies indicate that they are
quite unevolved.  They have low mass surface densities and this has
often been suggested as a possible cause for their slow evolution (van
der Hulst et al.\ 1987, McGaugh 1992, de Blok \& McGaugh 1996). 

Detailed investigations of a small sample of LSB galaxies (van der Hulst
et al.\ 1993) show that their gas surface densities in general lie below
the critical density needed for star formation, as derived by Kennicutt
(1989).  Although this global threshold density should be considered as
a boundary condition only (local instabilities may still cause star
formation), it nevertheless shows that conditions for star formation in
LSB galaxies are not as favourable as in ``normal'' high surface
brightness (HSB) galaxies. 

This might simply be caused by the low densities, making dynamical
timescales much longer, and therefore hampering the collapse of gas
complexes into Giant Molecular Clouds (GMCs).  The low metallicity may
also make cooling of the Interstellar Medium (ISM) more difficult,
delaying the formation of GMCs. 

To get a better handle on the properties of the cold, molecular
component of the ISM in LSB galaxies, one needs to observe indicators
such as the CO molecule.  Previous studies (Schombert et al. 1990)
that have tried to detect CO, have not succeeded to rather low limits.
This either means that CO does not work as a tracer in LSB galaxies --
implying that large amounts of molecular hydrogen could still exist --
or that LSB galaxies are deficient in their molecular component.

The case of H$_2$-poor galaxies is especially interesting: the
conditions that can then be deduced for LSB galaxies, which have
obviously formed stars, might help answer questions as: where and how
do stars form in an environment poor in molecular gas?  Is a small
molecular component, even as an intermediate agent always needed? Are
GMCs always needed for star formation?

In this paper we will describe the results and implications of a few
very deep pointed observations in the \hbox{CO(2-1)} line of various
galactic environments in LSB galaxies. The higher resolution of this
line with respect to the \hbox{CO(1-0)}-line enabled us to point at
different locations within one galaxy.  Positions were selected on the
basis of detailed H{\sc i} (de Blok et al.\ 1996) and optical imaging
(de Blok et al.\ 1995).  Section~2 describes the observations; in
sect.~3 the results are discussed; sect.~4 gives a discussion of the
implications; and sect.~5 summarizes the results. We will assume a
Hubble constant $H_0 = 75$ km s$^{-1}$ Mpc$^{-1}$ in the rest of this
paper.

\section{Observations}

Three galaxies were chosen from the sample of LSB galaxies described
in de Blok et al.\ (1996) and the references in the last paragraph of
the previous section. We refer to these papers for a description of
the detailed properties of our sample. In general, the sample contains
late-type LSB galaxies with absolute magnitudes $M_B \sim -17.5$,
central surface brightness $\mu_B \sim 23.5\ B$-mag arcsec$^{-2}$, and
colour $B-V \sim 0.5$.  In order to observe as wide a range of
galactic environments as possible, we used \HI column density maps and
H$\alpha$ imaging to choose prominent \HI minima and maxima and star
forming regions.  The H$\alpha$ images were also used to check whether
any \HI features coincided with optical features.  In the end four
positions were observed with the 15-m James Clerk Maxwell Telesope at
Mauna Kea, Hawaii, in the $^{12}$CO (J$=2-1$) line at 230 GHz
rest-frequency.

The observations were carried out from 29 March -- 3 April 1993.  The
A2 SIS receiver was used with the AOSC backend, giving 2048 channels
over a bandwidth of 500 MHz with a 250 kHz channel separation.  Because
of a factor of two oversampling the effective resolution was 500 kHz.
This corresponded to a velocity range of 652 km s$^{-1}$, and an
effective velocity resolution of 0.67 km s$^{-1}$ (2 channels).

The beam size was $22''$ and observations were made in beam-switching
mode.  The object position and a piece of sky 2 to 3$'$ away in
azimuth were observed.  Calibration was done with the help of a
three-load measurement.  Three resistors with known temperatures were
measured and thus calibrated the temperature scale.  These calibration
measurements were made every half hour.  The scatter in these
calibration measurements was less than 5\%  from night to
night.  This calibration was deemed to be sufficiently accurate for
our purpose.

The observed positions are given in Table 1, along with the name of
the galaxy, a description of the position and the total (on+off
source) integration time.  The top panels in Fig.\ 1 show
the pointing positions.

\begin{table}
\begin{minipage}{7.3cm}
\caption[]{Observed positions}
\begin{tabular}{lllccllccc}
\hline
Name &  $\alpha (1950.0)$ & $\delta (1950.0)$ & ${t_{int}}^a$ & type$^b$\\
\hline
F563-V1 &   08 43 46.7 & +19 04 16 & 8800  & peak\\
F568-V1 &   10 42 19.5 & +22 19 00 & 8800  & hole \\
        &   10 42 18.5 & +22 19 15 & 10855 & peak\\
F571-V1 &   11 23 40.5 & +19 06 50 & 10400 & SF \\
\hline
\end{tabular}
Notes: {\it a:} ${t_{int}}$ is total integration time in seconds. Time spent on
source is half this value.\\
{\it b:} `peak' denotes a peak in the \HI distribution, `hole' a
hole in the \HI distribution, `SF' a region of star formation.
\end{minipage}
\end{table}

\section{Results}

No CO emission was detected at any of the positions after on-source
integration times of $\sim 1.5$ hours per position.  Typical
RMS-noises at 500-kHz-resolution were $T_A^* \sim 6$ mK.  A beam efficieny
of 0.77 was used to convert the measured antenna temperatures $T_A^*$ to
brightness temperatures $T_b$.

Upper limits on the CO-flux were determined using a $3\sigma$ upper
limit.  As the original velocity resolution is too high to get any
meaningful upper limits, we have smoothed the spectra to lower
resolutions. We will present the results for two different
resolutions: a velocity channel separation of 5.2 km s$^{-1}$, i.e.,
identical to the velocity channel separation of the Schombert et al.\
(1990) observations, and 11 km s$^{-1}$, which is identical to the
velocity channel separation of the VLA \HI observations in de Blok et
al.\ (1996).

The $3\sigma$ upper limits to the H$_2$ mass in the beam were
determined following the method described in Schombert et al.\ (1990)
and Bregman \& Hogg (1988).  Using the \HI velocity widths measured
within the JCMT beam [extracted from the de Blok et al.\ (1996) \HI
data cubes], we can derive an upper limit to the H$_2$ mass as
follows.  For a channel spacing of 11 km s$^{-1}$ we get:

\[ I_{CO} \lid 3 T_b \cdot \Delta V_{HI} / \sqrt{n}, \]   
where $\Delta V_{HI} = 11 n$. This yields
\[
I_{CO} \lid 3 T_b  \cdot \Delta V_{HI} / \sqrt{\Delta V_{HI} / 11} = 
         3 T_b  \cdot \sqrt{11 \Delta V_{HI}} \]

We can then convert this to upper limits of H$_2$ masses in the
beam by using the formula given in Sanders et al.\ (1986):

\[ M(H_2) = 5.82 [ (\pi/4) d_b^2 I_{CO}].\] 

Here $d_b$ is the telescope beam diameter at the distance of the
source, expressed in parsecs.  This formula assumes $X = N(H_2)/ \int
T(CO) dv = 3.6\cdot 10^{20}$ cm$^{-2}$/(K\, km\, s$^{-1})$.  The H$_2$ mass depends directly on the
value of this conversion constant. In the next section we will show
that our conclusions do not depend crucially upon this factor.

\begin{figure*}
\epsfysize=0.45\vsize
\hfil\epsfbox{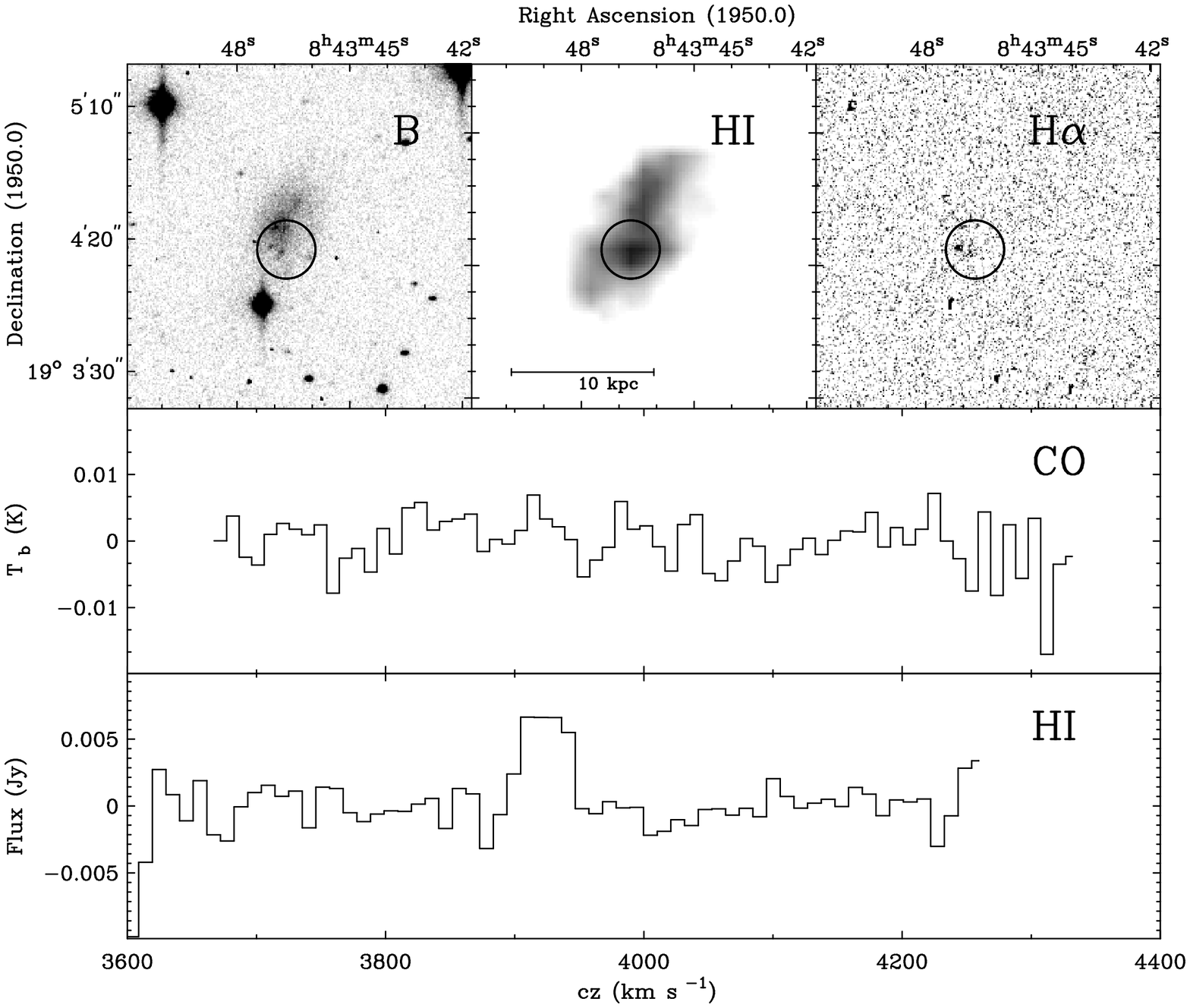}\hfil

\epsfysize=0.45\vsize
\hfil\epsfbox{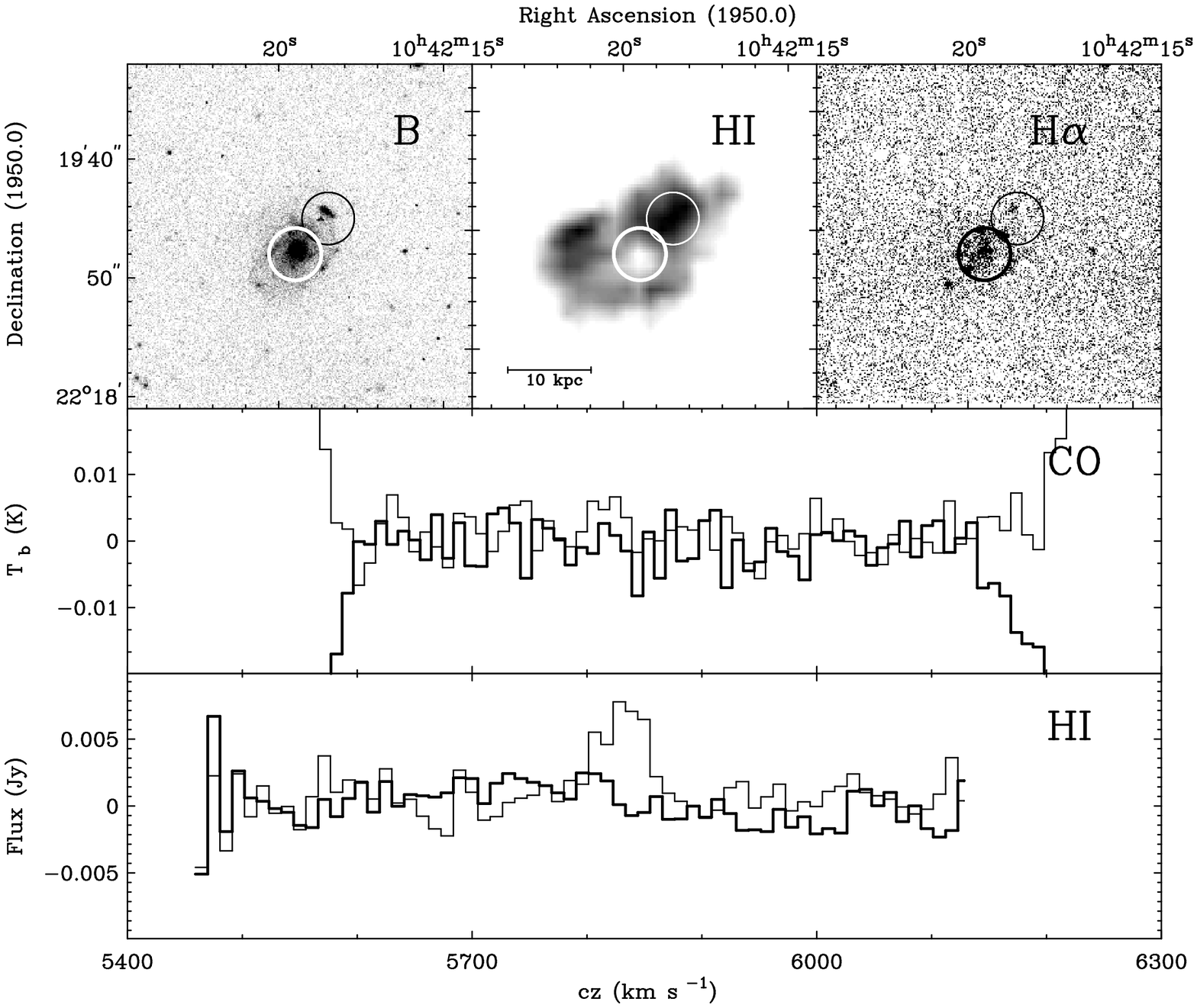}\hfil
\caption{Overview of the CO-observations of LSB galaxies. In each plot
  the top three panels show, from left-to-right, a $B$-band optical
  image, an \HI column density map and a continuum-subtracted
  H$\alpha$ image. Superimposed circles denote the position and size
  of the JCMT-beam. The panels in the center and bottom show the
  CO-spectrum and the \HI spectrum respectively, as measured at these
  positions. The top plot shows LSB galaxy F563-V1. The bottom plot
  LSB galaxy F568-V1. In this galaxy two positions were observed. The
  heavy lines denote position 1 (hole), the light lines position 2 (peak).}
\end{figure*}

\begin{figure*}
\epsfysize=0.45\vsize
\hfil\epsfbox{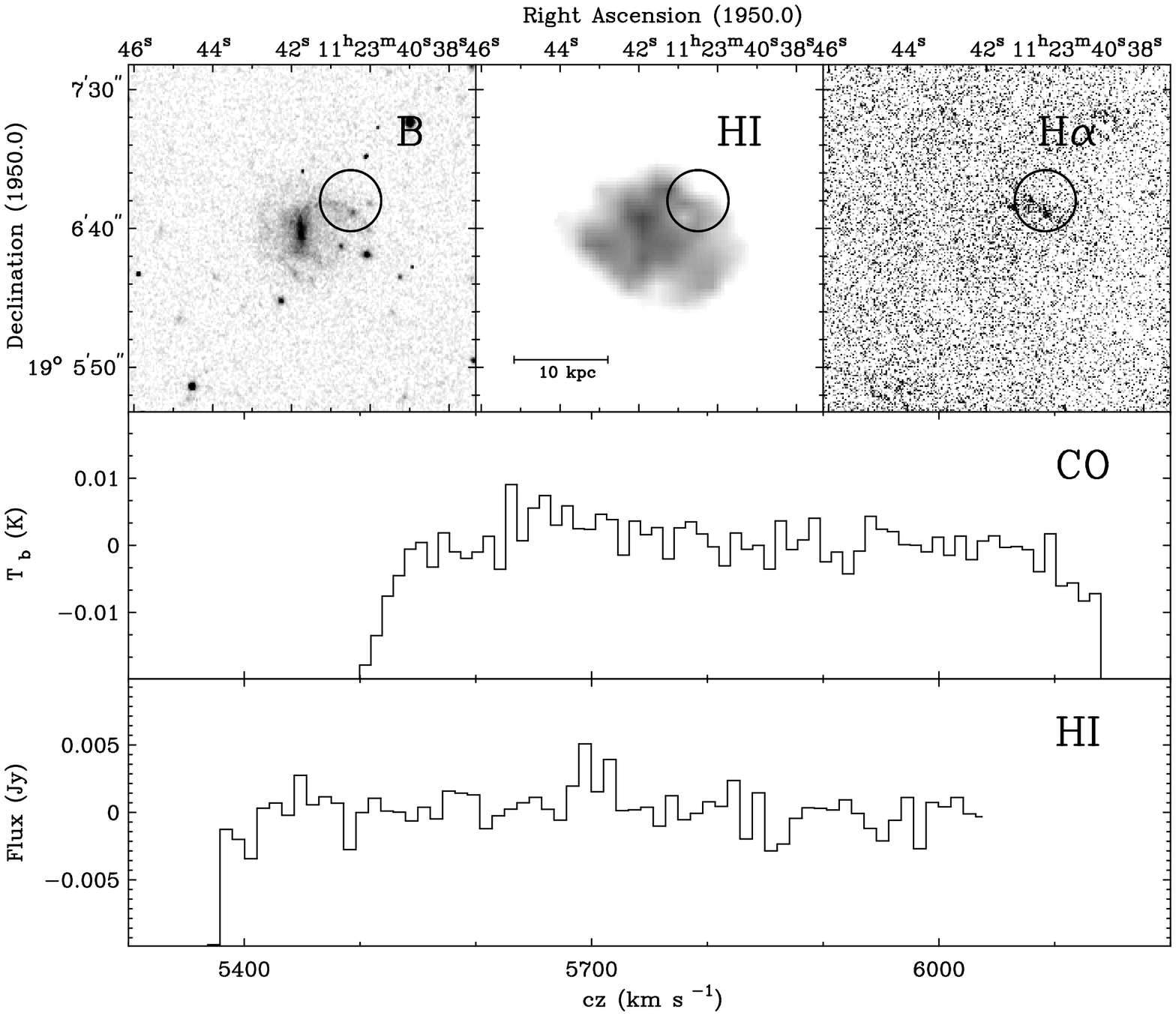}\hfil
\addtocounter{figure}{-1}
\caption{{\it (continued)} See previous caption. Shown is LSB galaxy F571-V1}
\end{figure*}

\begin{table*}
\begin{minipage}{17.3cm}
\caption[]{Upper limits to CO-fluxes and H$_2$ masses in LSB galaxies}
\begin{tabular}{lcllccllccc}
\noalign{\bf 5.2 km s$^{-1}$ resolution\strut}
\hline
\strut Name  &Pos   &  $\sigma(T_A^{*})$ &$\sigma(T_b)$ &$\Delta V_{HI}$&$I_{CO}$ &$D$ & $d_b$ & $M({\rm H}_2)$& $M({\rm H{\sc i}})$ &$M({\rm H}_2)/$\\

& & (mK) & (mK) & (km s$^{-1}$) & (K km s$^{-1})$ & (Mpc)
& (kpc) & ($10^7\ M_{\odot}$) &($10^7\ M_{\odot})$& $M({\rm H{\sc i}})$ \\
(1)&(2)&(3)&(4)&(5)&(6)&(7)&(8)&(9)&(10)&(11)\\
\hline
F563-V1&1&  3.76&   4.88&    65&   0.268&  51&     5.4&   3.57&  12.28  &0.28\\
F568-V1&1&  3.30&   4.28&    80&   0.260&  80&     8.5&   8.68&  20.09 &0.43\\
F568-V1&2&  2.40&   3.12&    80&   0.190&  80&     8.5&   6.31&  97.95& 0.064\\
F571-V1&1&  2.81&   3.64&    55&   0.185&  79&     8.4&   5.96&  24.00& 0.25\\
\hline
\noalign{\medskip}
\noalign{\bf 11 km s$^{-1}$ resolution\strut}
\hline
F563-V1&1&  2.77&   3.60&    65&   0.276&  51&     5.4&   3.63&  12.28& 0.30\\
F568-V1&1&  2.58&   3.35&    80&   0.188&  80&     8.5&   9.39&  20.09& 0.45\\
F568-V1&2&  2.20&   2.86&    80&   0.243&  80&     8.5&   8.04&  97.95& 0.081\\
F571-V1&1&  2.17&   2.82&    55&   0.198&  79&     8.4&   6.39&  24.00& 0.264\\
\hline
\end{tabular}
(1) Name of galaxy. (2) Position identification. (3) RMS noise in
antenna temperature. (4) RMS noise in brightness temperature. (5)
Velocity width \HI profile within JCMT beam. (6) Upper limit CO
intensity. (7) Distance to galaxy ($H_0=75$). (8) Diameter beam at
distance of galaxy. (9) 3$\sigma$ upper limit H$_2$ mass in beam. (10)
\HI mass in beam. (11) Upper limit mass ratio.\\

\end{minipage}
\end{table*}

Table 2 compares the RMS noises at the lower resolutions.  The top
panel contains the RMS-noise of the spectra smoothed to 5.2 km
s$^{-1}$; the bottom panel that of the 11 km s$^{-1}$ spectra.

The bottom panels of Fig.\ 1 compare the 11 km s$^{-1}$ JCMT spectra
with the VLA spectra measured at the same spatial position. These
latter were extracted from the data cubes using a beam size of
22\arcsec\ (the size of the JCMT beam).  If a significant amount of CO
were distributed like the \HI the CO profile should resemble the
neutral hydrogen profile at that pointing, but at the position of the
\HI signal, there is no hint of any CO emission.

The total masses of the \HI within the beam are compared with the
upper limits to the H$_2$ masses in Table 2.  The upper limits to the
$M$(H$_2$)/$M$(H{\sc i}) ratios are extremely low, consistent with the previous
measurements by Schombert et al. (1990) and Knezek (1993). While the
observations of Schombert et al. probed entire galaxies in one
observation, our pointed observations show that also more locally the
amount of CO is extremely low. 

Many studies list the total H$_2$ masses of galaxies and compare them
with other properties. To be able to compare our results, which only
give $M({\rm H}_2)$ in a part of the galaxy, with these other studies we
derive an upper limit on the total H$_2$ mass in the following way: we
compute the ratio between the area of the beam and the total area of
the galaxy within (1) the HI radius (radius where surface density
reaches 1 $M_{\odot}$ pc$^{-2}$) and (2) the optical radius $R_{25}$.
By multiplying the upper limits from Table~2 with this ratio we find
the total H$_2$ mass (again within either the HI radius or the optical
radius). We can then compare these with the total HI mass.  These
results are tabulated in Table~3. We have used the 11 km s$^{-1}$
results to be able to compare directly with the VLA data.

In general
the CO in other galaxies is found only at radii less than half of the
optical radius (Young \& Knezek 1989). In this respect the numbers
given in Table~3 are very optimistic numbers, as they assume that the
CO extends out to the HI radius and the optical radius,
respectively. The true numbers are therefore likely to be lower.
We will use the results derived for 11 km s$^{-1}$ and $R_{25}$ in the rest of this paper.

\begin{table*}
\begin{minipage}{135mm}
\caption[]{Upper limits to total H$_2$ masses}
\begin{tabular}{lrrrcrrc}
\noalign{\bf 11 km s$^{-1}$ resolution\strut}
\hline
\strut Name  &$M_{\rm HI}$   & $R_{\rm HI}$ & $M_{{\rm H}_2}$ & $M_{{\rm H}_2}/M_{\rm HI}$ &
$R_{25}$ & $M_{{\rm H}_2}$ & $M_{{\rm H}_2}/M_{\rm HI}$\\
   & ($10^7\ M_{\odot}$) & (kpc) & ($10^7\ M_{\odot}$) & 
   & (kpc) & ($10^7\ M_{\odot}$) & \\
\hline
F563-V1&  53.7&  6.5&  21.3&  0.393&   3.2&  5.1&  0.096\\
F568-V1& 245.5& 14.3& 106.5&  0.432&   6.2&  20.1&  0.081\\
F568-V1& 245.5& 14.3&  91.5&  0.372&   6.2&  17.1&  0.069\\
F571-V1& 117.5&  9.9&  35.4&  0.300&   3.8\rlap{$^a$}&  6.3& 0.054\\
\hline
\end{tabular}
{\it a:} $R_{25}$ is smaller than radius of beam. We have therefore used
the H$_2$ mass from Table~2 without changes. 
\end{minipage}
\end{table*}

\begin{figure}
\epsfxsize=\hsize\epsfbox{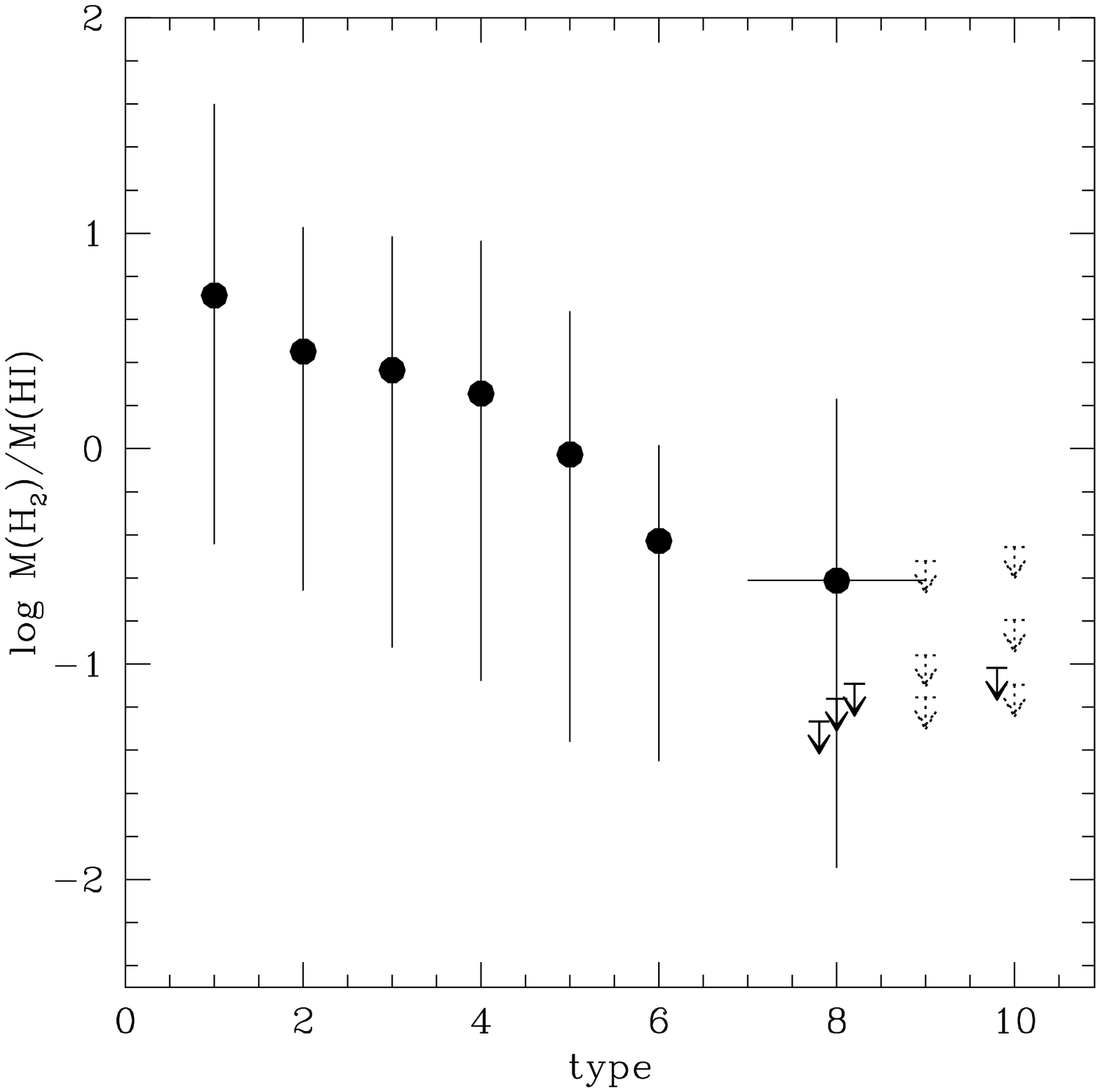}
\caption{The ratio of H$_2$ mass and \HI\ mass plotted as function of
  Hubble type. The filled circles show the average value for each
  Hubble type from Young \& Knezek, indicating a decreasing importance
  of the molecular component with respect to the atomic component. The
  vertical lines indicate the {\it full range} of the Young \& Knezek
  data points. The Schombert et al.\ measurements are indicated by the
  dotted symbols. Our measurements are indicated by the solid upper limit
  symbols. The data assume a constant conversion factor of $3.6 \times
  10^{20}$  cm$^{-2}$/(K\, km\, s$^{-1})$. }
\end{figure}

\begin{figure}
  \epsfxsize=\hsize \epsfbox{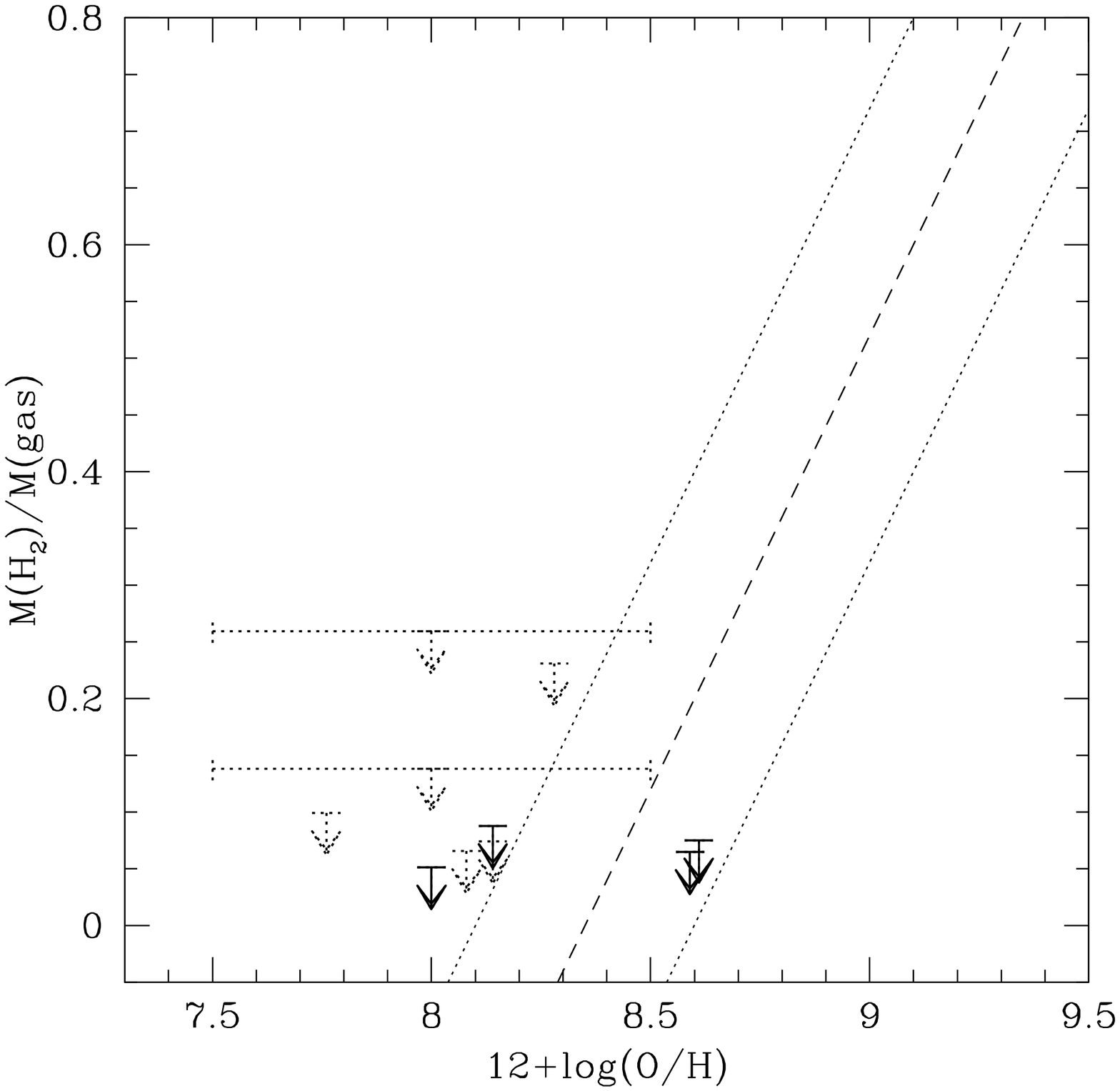}
\caption{The ratio of H$_2$ mass and total gas mass plotted versus the
  oxygen abundance. The dashed line show the average relation derived
  for HSB galaxies by VCE, the dotted lines show the approximate
  2$\sigma$ width of the distribution. Our measurements are indicated by the solid symbols, while  those of
  Schombert et al. (1990) are indicated by the dotted upper limit symbols.
  We used abundance data from McGaugh (1994) and Paper I. Two cases
  where no abundance measurements were available are indicated by the
  horizontal errorbars that span the range of abundances found in LSB
  galaxies. The upper limits are consistent with the trend derived by
  VCE.}
\end{figure}

\section{Discussion}

The non-detections of CO can be taken at face-value to suggest that LSB
galaxies are poor in H$_2$.  There are, however, several factors which
complicate this naive interpretation.  The most important of these is
the conversion factor $X$ which is used to convert the measured CO
brightness temperature into an H$_2$ mass.  The value of $X$ is
uncertain and is inferred to have a large range.  Any interpretation of
CO measurements will thus depend on the assumed values of $X$. 

In the following sections we will first discuss our results in the light
of other studies of the molecular gas in late type galaxies, then
explore the implications of possible variations in the conversion factor
$X$ with morphological type and metallicity and finally infer that the
LSB galaxies of the kind considered are here significantly poorer in H$_2$
than their HSB counterparts. 

\subsection{Comparison with other galaxies} 

As the ISM in LSB galaxies has  low metallicities (McGaugh 1994) we
will compare our non-detections with observations of samples of other
late-type, low-metallicity galaxies.

One such sample is that of Sage et al.\ (1992) of dwarf irregulars and
blue compact galaxies.  One of their conclusions is that for the
galaxies in their sample the CO/\HI ratio did not depend on
metallicity.  If this were also true for our LSB galaxies then 
we would have expected to detect CO based on their \HI masses.

Our results therefore appear inconsistent with a constant CO/\HI ratio. 
This should perhaps not be expected: dwarf irregulars and especially
blue compact galaxies have appreciable star formation rates or are
undergoing bursts and both also have high surface densities.  LSB galaxies
on the other hand have low current star formation rates and low \HI
surface densities. 

A better comparison sample therefore is perhaps the sample of Magellanic
irregular galaxies of Hunter \& Sage (1993).  These observations also
resulted in null-detections.  Based on this they suggest that molecular
gas may be a transient phenomenon in dwarf galaxies, as a result of the
low \HI volume densities which supposedly are not high enough to sustain
H$_2$.  The low \HI densities found in LSB galaxies (de Blok et al.\
1996) then suggest that similar reasoning can explain the apparent lack
of CO, and thus of H$_2$. 

It is of interest also to examine as a comparison sample the sample of
late-type galaxies in the Virgo cluster as described by Kenney \& Young
(1988).  They discuss several reasons why the CO-poor but H{\sc i}-rich
galaxies in their sample are most likely also H$_2$-poor.  The LSB
galaxies discussed here share some of the properties of the late-type
Virgo-spirals so 
this discussion is relevant for our results.

The sample of Virgo spirals shows a large range in CO-surface
brightness (defined as $L_{CO}/D_{opt}^2$) with the most luminous
galaxies having the highest values. Kenney \& Young divide the group
of galaxies with $M_B \sim -19$ into two groups according to their
H{\sc i}-richness. The H{\sc i}-rich galaxies turn out to be poor in
CO, while the H{\sc i}-poor galaxies are rich in CO. As they find no
H{\sc i}-rich galaxies that are also CO-rich, they suggest that this
indicates that the H{\sc i}-rich galaxies do not contain much
molecular gas. This might thus be a hint that the gas-rich late-type
LSB galaxies are also poor in molecular gas.

Kenney \& Young also find a correlation between the amount of far
infra-red (FIR) emission (arising from dust heated by starlight) and
the total amount of gas within the optical radius of a galaxy.  Actual
FIR measurements of LSB galaxies are to our knowledge not available
(None of the late-type LSB galaxies discussed here was found in the
IRAS database), but we can use the work by Bothun et al.\
(1989) to show that LSB galaxies must have FIR luminosities of $\la
10^8\ L_{\odot}$. Comparison with the relation of Kenney \& Young
(their Fig.\ 3d) then implies that the amount of \HI present in LSB
galaxies (a few times $10^9\ M_{\odot}$) is (more than) enough to
account for the total gas mass. This also is an indication that LSB
galaxies do not contain large amounts of molecular gas.

\subsection{The CO -- H$_2$ conversion factor $X$}

What conditions must a galaxy fulfill in order to be able to maintain
large amounts of molecular gas? First, the metallicity must be high
enough to ensure sufficient amounts of dust.  Second, the total column
density of gas must be large enough to shield H$_2$ from dissociating
radiation. Measurements of colours, metallicities and H{\sc ii} region
Balmer decrements of late-type LSB galaxies all suggests a low dust
content (McGaugh 1994, van den Hoek et al.\ 1997). Observations of the
\HI (de Blok et al. 1996) show low surface densities.  LSB galaxies do
not seem to obey the conditions necessary for H$_2$.

As the H$_2$ formation rate is thought to be proportional to the number
density of the dust, the low dust content and low dust-to-gas ratio,
or alternatively the low metallicity, lower this rate and also lower
the fraction of star-forming-cloud mass that is in molecular form.
The low dust content may make the conversion from \HI into H$_2$ more
difficult, as dust grains provide shielding from the interstellar
radiation field.  A larger column density of gas is thus needed to
self-shield the H$_2$. The low densities and lack of dust found in LSB
galaxies make it easier for UV photons to dissociate molecules, thus
destroying H$_2$ more easily.

However, the same processes affect the CO molecules even more.  Lower
abundances of oxygen, and presumably carbon, in the gas means that the
sizes of the clouds as traced by the CO may be smaller than the
underlying H$_2$ clouds. The CO emission will therefore be lower too.
If a constant conversion factor is assumed, this then leads to an
underestimate of the total amount of molecular hydrogen.

This effect is dramatically illustrated by Maloney \& Black (1988)
using a model of the SMC.  A dust to gas ratio (equivalent to
metallicity) 17 times smaller than that of the local solar neigbourhood
yields an H$_2$ peak abundance which is decreased by only 10\%. 
The effect is much more dramatic for the CO.  Whereas models for
Galactic GMCs show that 99\% of the carbon is locked up in CO,
the GMCs in the SMC contain only 1\% of the carbon in the form of
CO.  Self-shielding is thus very important for the survival of CO
molecules in the interstellar UV radiation field.  Observations of
molecular clouds in the SMC (Rubio et al.\ 1993) do indeed show this
effect and the value found for $X$ from an analysis of virial masses of
individual clouds is 4 -- 20 times higher than the standard Galactic
value for clouds of 20 -- 200 pc in size. 

Similarly the CO to H$_2$ conversion factor can be very
different in the low metallicity LSB galaxies. One should keep in
mind that a different value of $X$ {\it does not imply that LSB
galaxies contain large amounts of \rm H$_2$.} It still leaves open the
possibility that the absolute amount of H$_2$ may be fairly small.
We therefore discuss various approaches to the conversion factor
and the implications for the H$_2$ content of LSB galaxies in the
following subsections (4.2.1-3)

\subsubsection{A constant conversion factor}

Young \& Knezek (1989) have analysed the change in $M({\rm
H}_2)/(M_{\rm HI})$ over a large range in Hubble type, and found a
decreasing importance of H$_2$ towards later Hubble types.  In this
analysis they kept the conversion factor $X$ constant at $2.8 \times
10^{20}$ cm$^{-2}$/(K\, km\, s$^{-1})$, arguing that this trend cannot
be caused by a change in $X$ alone, as that would imply that the
temperature of the gaseous ISM would need to be some 20 times lower in
late type galaxies than in early types, or alternatively the density
of the gas would have to differ by a factor of 400 between early and
late types.

This means that the change in $M({\rm H}_2)/M({\rm H{\sc i}})$ would
have to be caused at least partly by a true decrease in the importance
of H$_2$ towards later types.  Again, if these arguments are accepted,
LSB galaxies are galaxies that are poor in H$_2$, and whose gas
component is totally dominated by the neutral hydrogen.

This is illustrated in Fig.\ 2, where the data of Young \& Knezek is
schematically represented, along with LSB measurements.  The upper
limits derived for the LSB galaxies clearly follow the trend defined
by the Hubble sequence.  Assuming a constant $X$ thus makes LSB
galaxies 2 orders of magnitude more poor in H$_2$ than early-type
HSB galaxies 
and, based on our upper limits, a factor of $\sim 5$ poorer than HSB
late-type galaxies of similar Hubble type.

\subsubsection{An ad-hoc conversion factor}

Low H$_2$ fractions can also be inferred from the work of Vila-Costas
\& Edmunds (1992) [VCE].  Amongst other things they looked at the
mutual dependences in a number of HSB galaxies of metal abundances in
the gas, surface densities of the gas and H$_2$ fractions.  The latter
were determined from CO fluxes from the literature, by assuming a
variable conversion factor that would for each galaxy give an
exponential total (i.e.\ \HI and H$_2$) gas distribution.  They show
that low gas-surface density galaxies have low oxygen abundances (their
Fig.\ 7), and low H$_2$ fractions. 

In Fig.\ 3 the relation between abundance and molecular gas fraction
from VCE is shown, with our measurements and those from Schombert et
al.\ (1990) overplotted.  We did not include Knezek's (1993)
sample as this consists of giant LSB galaxies who have most
likely undergone a different evolutionary history.

Applying the abundance values from McGaugh (1994) and de Blok \& van
der Hulst (1997, Paper I) to the trend derived by VCE we find that
$M({\rm H}_2)/M({\rm gas})$
in LSB galaxies must be less than $0.25$, consistent
with the actual measurements, where our measurements imply ratios of
less than 0.06.

The trend as presented in VCE cannot be explained as an artefact of
their variable conversion factor $X$.  VCE allow values between 0.8
and 4.8 ($\times 10^{20}$), which is a factor of 6.  This translates
in approximately a factor of 2 change in $M({\rm H}_2)/M({\rm gas})$.  The various
values of $X$ which VCE derive are however independent of properties
like Hubble type, and will therefore not introduce any {\it
systematic} trends.  The effect of changing $X$ in a non-systematic
way as VCE have done is therefore merely a shift in the positions of
individual galaxies by at most a factor of 2.  In essence they have
just added scatter to the relation derived by Young \& Knezek (1989).
This therefore does not affect the conclusion that the VCE analysis
implies that the molecular component in LSB galaxies most likely
constitutes only a small fraction of the total amount of gas.

\subsubsection{A metallicity-dependent conversion factor}

Wilson (1995) demonstrated that Maloney \& Black's (1988) ideas
concerning a variation in $X$ with metallicity is borne out in
observations of galaxies in the Local Group.  Based on measurements of
the CO luminosity and determination of the virial masses of individual
clouds,  Wilson finds that the conversion factor increases as the
metallicity decreases.  Israel (1997) investigated the metallicity
dependence of $X$ in a different way using the FIR surface brightness
and \HI column density to estimate the column density of H$_2$ and found
an even steeper relationship.  

The average oxygen abundance for the LSB galaxies we observed is 12 +
log(O/H) $\sim 8.4$.  Using the above results this would lead to
conversion factor values $X$ of 2 -- 6 times the Galactic value. 

The star formation rates and \HI column densities in the galaxies used
to derive the dependence of $X$ on metallicity are appreciably higher
than those commonly found in LSB galaxies. The lower star formation rate
implies a lower energy density of the radiation field and consequently
lower dissociation of the CO and H$_2$. The result will be that $X$
probably is not as large as in the extreme case of the SMC, so some
care should be excercised in using these results for estimating the
H$_2$ mass limits for LSB galaxies. 

Another effect of the low metallicities is a less efficient cooling of
the ISM, which leads to {\it higher} cloud temperatures, making it
difficult for a cold molecular phase to exist.  Detailed modelling
suggests that the lack of cooling is sufficient to prevent most of the
gas from becoming cold ($T<1000\ K$) (see the results presented in
Gerritsen \& de Blok [Paper III]), and that this is one of the main
causes for the low star formation rates in LSB galaxies.  This implies
that H$_2$ would only be a small fraction of the total gas mass.

Bearing these effects in mind we estimate that $X$ will be $\sim 4$
times the Galactic value in our objects.  In other words, LSB galaxies
should contain 4 times more H$_2$ than the Galactic value suggests. 

The trend found by Young \& Knezek (1989) implies a decrease in the
importance of molecular gas by a factor of 300 from early- to
late-type galaxies.  Our LSB galaxies are furthermore at least a
factor 10 more poor in molecular gas than average late-type HSB
galaxies.

Assuming that the Galactic value of $X$ holds for the early-type
galaxies, the metallicity dependence of $X$ increases the upper limits
for the amount of molecular gas inferred in LSB galaxies by a factor
of $\sim 4$.  This still makes LSB galaxies a factor $\sim 75$ more
poor in molecular gas than the early-types.  The metallicity
dependence of $X$ implies that its value for the late-type HSB
galaxies will also be larger. This means that the amount of molecular
gas in LSB galaxies will increase by {\it less than} a factor of $\sim
4$ with respect to the late-type HSB galaxies, thus retaining the
difference between LSB and HSB galaxies. If, as suggested by, amongst
others, Israel (1997) the value of $X$ depends mainly on the radiation
field, rather than metallicity, then the value of $X$ in late-type HSB
galaxies could actually be $\it higher$ than in the LSB galaxies, thus
increasing the difference between HSB and LSB galaxies.

Similar arguments apply to the VCE results.  The larger $X$ will
decrease the slope of the VCE trend by about a factor of 3, which is
however not enough to make the trend of decreasing H$_2$ fraction with
decreasing abundance disappear.

In summary, the metallicity dependence of $X$ tends to offset partly
the trends found by Young \& Knezek (1989) and VCE, but the effect is
not strong enough to make these trends disappear.  The trend of
decreasing molecular gas content with Hubble type remains, although
slightly less steep than given by Young \& Knezek.  The conclusion
remains that it is likely that LSB galaxies have smaller H$_2$
fractions than their HSB counterparts.
 
\subsubsection{Other arguments}

An additional argument why it is plausible to have small H$_2$
fractions in LSB galaxies is the role of shear. One way of creating
high column density regions where molecular gas may form is by making
massive clouds. These form most likely in cloud-cloud collisions. The
collision rate will be larger in galaxies where shear plays an
important role. Clouds may then also grow from gravitational accretion
in shearing gas layers.  The rotation curves of LSB galaxies show them
to have only slowly rising rotation curves with large solid-body
parts, so that the amount of shear will be smaller with a consequently
smaller cloud growth rate.

The cloud formation rate also depends on the mean gas volume
density. For example in our Milky Way at 4 kpc 70 percent of the
molecular gas is locked up in massive clouds, while at 10 kpc this is
only 10 percent (see e.g. Sakamoto et al.\ 1997).  It will be clear
that the conditions for forming massive clouds will be less favourable
in LSB galaxies.

These arguments thus imply that LSB galaxies probably have a low H$_2$
content.  We should note though that CO emission has been detected in
a few LSB galaxies in the sample of Knezek (1993). However, as noted
earlier, her sample was selected to contain giant early-type LSB
galaxies. These galaxies have a much different morphology
(e.g. presence of a large bulge) than the galaxies in our sample. A
detailed discussion of these galaxies would be interesting but is
unfortunately beyond the scope of this paper.

\subsection{Caveats}

Throughout this paper we have assumed a one-to-one correspondence
between the CO(2-1) brightness temperature and the CO(1-0) brightness
temperature. All derivations of $X$ etc. are based on the latter.  In
practice the correspondence is not entirely one-to-one, as the CO(1-0)
and CO(2-1) lines do not necessarily trace the same gas. CO(2-1)
probably traces slightly warmer and denser clouds.  Chiar et
al. (1994) find from observations of an ensemble of molecular clouds
in the Galactic plane that the average ratio between the CO(2-1) and
CO(1-0) brightness temperatures $T_{2-1}/T_{1-0}$ is 0.8.  This
implies that a conversion factor $X_{2-1}$ based on CO(2-1)
observations should be 20 percent larger than $X_{1-0}$, which is the
commonly used conversion factor.

Further independent modelling by Kutner et al.\ (1990) shows
that the ratio $T_{2-1}/T_{1-0}$ depends on gas density, CO abundance
and temperature.  They find that $T_{2-1}/T_{1-0}$ increases with
decreasing CO abundance.  One might tentatively conclude from this that
the $T_{2-1}/T_{1-0}$ ratio in LSB galaxies might be closer to unity
than the value of 0.8 mentioned above.  The results of Rubio et al.\
(1993) for the SMC indicate a $T_{2-1}/T_{1-0}$ ratio of 1.2, supporting
the idea that this ratio is at least unity for LSB galaxies.  In the
worst case assuming a one-to-one correspondence between CO(2-1) and
CO(1-0) (i.e.  assuming that $T_{2-1}/T_{1-0} = 1$) thus underestimates
the derived LSB H$_2$ masses by 20 percent.  This is not enough to make
LSB galaxies rich in H$_2$ and does affect neither the discussion nor
the results.

\section{Concluding Remarks}

We have presented deep, pointed $^{12}$CO($J=2-1$) observations of
three LSB galaxies. No CO was found at any of the positions observed.
This leads to a mean upper limit of the local $M({\rm H}_2)/M({\rm
H{\sc i}}) < 0.25$, with individual values for the total $M({\rm
H}_2)/M({\rm H{\sc i}})$ ratio reaching less than 6\% (assuming a
Galactic value for the CO to H$_2$ conversion factor $X$).

It is, however, unlikely that the Galactic conversion factor applies
to LSB galaxies which have low metallicities and low dust content. In
fact the value for $X$ is likely to be $\sim 4$ times higher. The
H$_2$ content would be correspondingly higher.  Our limits then imply
that LSB galaxies roughly have (less than) 25\% of their gass mass in
the form of H$_2$.  This is still lower than is found in HSB galaxies.

The conclusion then is that there are no large amounts of H$_2$ hidden
in LSB galaxies.  The low star formation rates measured in LSB
galaxies can thus be explained by the virtual absence of a molecular
component. Star formation in LSB galaxies may thus proceed in a
different way than in HSB galaxies.  A detailed comparison between the
properties of star forming regions in LSB and HSB galaxies may be a
good way to put more constraints on the way stars form in environments
that lack a cold component.

\begin{acknowledgements}
We thank the anonymous referee and Tom Oosterloo for comments that
have improved aspects of this paper.

The James Clerk Maxwell Telescope is operated by The Joint Astronomy
Centre on behalf of the Particle Physics and Astronomy Research
Council of the United Kingdom, the Netherlands Organisation for
Scientific Research, and the National Research Council of Canada.
\end{acknowledgements}

\end{document}